\documentclass[twocolumn,showpacs,aps,prd]{revtex4}
\pdfoutput=1
\newcommand{\half}{\frac{1}{2}}
\newcommand{\thrd}{\frac{1}{3}}
\newcommand{\frth}{\frac{1}{4}}
\newcommand{\sxth}{\frac{1}{6}}
\newcommand{\dar}{\partial_r}
\newcommand{\R}{{\bar r}}
\usepackage[dvipsnames]{color}
\newcommand{\revision}[1]{{            {#1}}} 
\begin{document}
\title{Implications of the conformal Higgs model}
\author{R. K. Nesbet }
\affiliation{
IBM Almaden Research Center,
650 Harry Road,
San Jose, CA 95120-6099, USA
\begin{center}rkn@earthlink.net\end{center}
}
 \date{\today}
\begin{abstract}
The postulate of universal local Weyl scaling (conformal) symmetry 
modifies both general relativity and the Higgs scalar field model.
The conformal Higgs model (CHM) acquires a cosmological effect that 
fits observed accelerating Hubble expansion for redshifts $z\leq 1$ 
(7.33 Gyr) accurately with only one free constant parameter.  Conformal 
gravity (CG) has recently been fitted to anomalous rotation data for 
138 galaxies. Conformal theory explains dark energy and does not 
require dark matter, providing a viable alternative to the $\Lambda$CDM 
standard paradigm.  The theory precludes a massive Higgs particle but
validates a composite gauge field $W_2$ with mass 125GeV.
\end{abstract}
\pacs{04.20.Cv,98.80.-k,11.15.-q} 
\maketitle
\section{Introduction}
\par Gravitational phenomena that cannot be explained by general 
relativity as formulated by Einstein are attributed to cold dark matter
in the consensus $\Lambda$CDM paradigm for cosmology.  The search for 
tangible dark matter has continued without result for many 
years\cite{SAN10}. Dark energy $\Lambda$ remains without explanation.
\par Universal conformal symmetry offers an alternative paradigm, 
motivated by this situation. This requires local Weyl scaling covariance 
\cite{WEY18,MAN06,MAN07} for all massless elementary physical fields, 
without dark matter\cite{NES13}. With no novel elementary fields,
this extends conformal symmetry, valid for fermion and gauge boson 
fields\cite{DEW64}, to both the metric tensor field of general 
relativity and the Higgs scalar field of elementary-particle 
theory \cite{HIG64,CAG98}.  This postulate implies conformal
gravity CG\cite{MAK89,MAK94,MAN90,MAN91,MAN06,MAN12} and 
the conformal Higgs model CHM\cite{NESM1,NESM2,NESM3,NESM5}. 
\par The CHM determines model parameters that retain the Higgs 
mechanism but preclude a massive Higgs particle, replacing mass 
by implied dark energy in agreement with observed Hubble 
expansion \cite{NESM1,NESM2}, while validating a novel composite
gauge field $W_2$ of mass 125GeV \cite{NESM4}.  This paper reviews
the logic of the CHM and its qualitative and quantitative justification 
by observed cosmological, gravitational, and particle physics data.
\par Conformal gravity (CG)\cite{MAN06} retains the logical structure
of general relativity, replacing the Einstein-Hilbert Lagrangian
density by a quadratic contraction of the conformal Weyl
tensor\cite{WEY18}.  CG ensures consistency of the
gravitational field equations, while preserving subgalactic
phenomenology\cite{MAK89,MAN06}.  The conformal Higgs model (CHM)
introduces a gravitational term confirmed by observed 
accelerating Hubble expansion\cite{MAN06,NESM1,NESM2,NESM5}.
\par Substantial empirical support for this proposed break with
convention is provided by recent applications of CG to 
galactic rotation velocities\cite{MAN97,OCM18,MAO11,MAO12,OAM12,OAM15}
and of the CHM to Hubble expansion \cite{NESM1,NESM2}, in a consistent 
model of extended dark galactic halos\cite{NESM3}, 
\revision{
without dark matter. Gravitational lensing by a galactic halo is 
attributed to the boundary condition of continuous gravitational
acceleration across the outer boundary radius of the large spherical
halo depleted of primordial matter fallen into the visible
galaxy\cite{NESM3}.  This accounts for the nonclassical acceleration
parameter $\gamma$ deduced from anomalous orbital rotation velocities
observed in the outer regions of galaxies\cite{MAN06}.
}
As recently reviewed \cite{MAN06,MAN12,NESM5}, conformal theory fits 
observed data for an 
isolated galaxy without invoking dark matter, resolving several 
longstanding paradoxes.  Higgs biquadratic parameter $\lambda$ is 
found to determine $W_2$ mass 125GeV \cite{NESM4}. Derivations of these 
results are summarized here. 

\section{Depleted halo model}
\par CG and the CHM are consistent but interdependent\cite{NESM5} 
in the context of a depleted dark halo model \cite{NESM3} of an 
isolated galaxy.  A galaxy of mass $M$ is modeled by spherically
averaged mass density $\rho_G/c^2$ within $r_G$, formed by
condensation of primordial uniform, isotropic matter of uniform mass
density $\rho_m/c^2$ from a sphere of large radius $r_H$ \cite{NESM3}.
The dark halo inferred from gravitational lensing and centripetal
acceleration is identified with this depleted sphere \cite{NESM3}.
A model valid for nonclassical gravitation can take advantage of
spherical symmetry at large galactic radii, assuming classical
gravitation within an effective galactic radius $r_G$.  Non-spherical
gravitation is neglected outside $r_G$. Given mean mass density
${\bar\rho}_G/c^2$ within $r_G$, this implies large empty halo radius
$r_H=r_G({\bar\rho}_G/\rho_m)^\thrd$.
\par The unique Lagrangian density ${\cal L}_g$ of conformal gravity
theory, constructed from the conformal Weyl
tensor \cite{WEY18,MAK89,MAN06}, determines source-free
Schwarzschild gravitational potential
\begin{eqnarray} \label{Brfn}
B(r)=-2\beta/r+\alpha+\gamma r-\kappa r^2,
\end{eqnarray}
valid outside a spherically symmetric mass/energy source density
\cite{MAK89,MAN91}. This adds two constants of integration to the 
classical external potential: nonclassical radial acceleration 
$\gamma$ and halo cutoff parameter $\kappa$ \cite{NESM3}. 
\par The physically relevant particular solution for $B(r)$\cite{NESM5}
incorporates nonclassical radial acceleration $\gamma$ as a free 
parameter.  Its value is determined by the halo model.  
Gravitational lensing by a spherical halo is observed as centripetal
deflection of a photon geodesic passing into the empty halo sphere from
the external intergalactic space with postulated universal isotropic
mass-energy density $\rho_m$.  The conformal Friedmann cosmic evolution
equation implies dimensionless cosmic acceleration parameters 
$\Omega_q(\rho)$\cite{NESM3} which are locally constant but differ 
across the halo boundary $r_H$.  Smooth evolution of the cosmos implies 
observable particle acceleration $\gamma$ within $r_H$ proportional to 
$\Omega_q(in)-\Omega_q(out)=\Omega_q(0)-\Omega_q(\rho_m)$.  Uniform 
cosmological $\rho_m$ implies constant $\gamma$ for $r\leq r_H$,
independent of galactic mass \cite{NESM5}.  This surprising result is 
consistent with recent observations of galactic rotational velocities
for galaxies with directly measured mass\cite{MLS16,NESM7}.

\section{Variational field equations}
\par Gravitational
field equations are determined by metric functional derivative
$X^{\mu\nu}= \frac{1}{\sqrt{-g}}\frac{\delta I}{\delta g_{\mu\nu}}$,
where $g$ is the determinant of $g_{\mu\nu}$, and action integral
$I=\int d^4x \sqrt{-g}\sum_a{\cal L}_a$ \cite{WEI72,NES03,MAN06}.
Given $\delta{\cal L}=x^{\mu\nu}\delta g_{\mu\nu}$,
$X^{\mu\nu}=x^{\mu\nu}+\half{\cal L}g^{\mu\nu}$.
Scalar Lagrangian density ${\cal L}_a$ determines variational
energy-momentum tensor $\Theta_a^{\mu\nu}=-2X_a^{\mu\nu}$, evaluated
for a solution of the field equations.
\par For fixed coordinates $x^\mu$, local Weyl scaling is defined by
$g_{\mu\nu}(x)\to g_{\mu\nu}(x)\Omega^2(x)$\cite{WEY18} for arbitrary
real differentiable $\Omega(x)$. Conformal symmetry is defined by
invariant action integral $I=\int d^4x\sqrt{-g}{\cal L}$.
For any Riemannian tensor $T(x)$, $T(x)\to \Omega^d(x)T(x)+{\cal
R}(x)$ defines weight $d[T]$ and residue ${\cal R}[T]$. For a scalar
field, $\Phi(x)\to\Phi(x)\Omega^{-1}(x)$, so that $d[\Phi]=-1$.
Conformal Lagrangian density ${\cal L}$ must have weight
$d[{\cal L}]=-4$ and residue ${\cal R}[{\cal L}]=0$ up
to a 4-divergence\cite{MAN06}.  For a bare conformal
field, trace $g_{\mu\nu}X_a^{\mu\nu}=0$ \cite{MAN06}.  
Generalized Einstein equation $\sum_aX_a^{\mu\nu}=0$ is expressed as
$X_g^{\mu\nu}=\half\sum_{a\neq g}\Theta_a^{\mu\nu}$.  Summed trace
$\sum_ag_{\mu\nu}X_a^{\mu\nu}$ vanishes for exact field solutions. 
\par The action integral defined by Lagrangian density
${\cal L}_W=-\alpha_g C_\lambda^{\mu\kappa\nu}C^\lambda_{\mu\kappa\nu}$
for Weyl tensor $C_\lambda^{\mu\kappa\nu}$, a traceless projection of
the Riemann tensor \cite{WEY18}, is conformally invariant.  
After removing a 4-divergence \cite{MAN06},
${\cal L}_g=-2\alpha_g( R^{\mu\nu}R_{\mu\nu}-\thrd R^2 )$, where
$R=g_{\mu\nu}R^{\mu\nu}$.  Conformal
symmetry fixes the relative coefficient of the two quadratic terms. 
For uniform mass/energy density ${\bar\rho}$ the Weyl tensor vanishes 
identically. Hence $X_g^{\mu\nu}\equiv 0$ for a uniform cosmos. 

\section{The conformal Higgs model}
\par Higgs $V(\Phi^\dag\Phi)=-(w^2-\lambda\Phi^\dag\Phi)\Phi^\dag\Phi$
depends on two assumed constants $w^2$ and $\lambda$\cite{HIG64,CAG98}.
Nonzero $w^2$ and $\lambda$ are not determined by standard theory. 
The conformal Higgs model (CHM) introduces a gravitational term 
confirmed by observed Hubble expansion\cite{NESM1,NESM2}.  The CHM 
supports the Higgs mechanism, spontaneous SU(2) symmetry-breaking, 
which also breaks conformal symmetry\cite{NESM2} and invalidates a 
transformation connecting the two distinct metrics invoked for 
successful fits to galactic rotation and Hubble expansion\cite{MAN06}.
\par In the conformal Higgs model, unique Lagrangian density
${\cal L}_\Phi$\cite{MAN06,NESM1,NESM2} of Higgs scalar field
$\Phi$ adds a gravitational term to $-V$, so that
$(\partial_\mu\Phi)^\dag\partial^\mu\Phi$ is augmented by
\begin{eqnarray} \label{DeltaL}
\Delta{\cal L}_\Phi=
 (w^2-\sxth R-\lambda\Phi^\dag\Phi)\Phi^\dag\Phi.
\end{eqnarray}
Scalar $R=g_{\mu\nu}R^{\mu\nu}$ is the trace of the Ricci tensor.
\par In uniform, isotropic geometry with uniform mass/energy density
${\bar\rho}$, ${\cal L}_\Phi$
implies a modified Friedmann equation 
\cite{FRI22,NESM1,NES13} for cosmic distance scale factor $a(t)$,
with $a(t_0)=1$ at present time $t_0$:
\begin{eqnarray} \label{MFeq}
 \frac{{\dot a}^2}{a^2}+\frac{k}{a^2}-\frac{\ddot a}{a}
    =\frac{2}{3}({\bar\Lambda}+{\bar\tau}c^2{\bar\rho}(t)).
\end{eqnarray}
${\bar\Lambda}=\frac{3}{2}w^2\geq 0$ and 
${\bar\tau}\sim-3/\phi_0^2\leq 0$ are determined by 
parameters of the Higgs model\cite{NESM1}.
\par Neglecting cosmic curvature $k$ in accord with observed data,
sum rule $\Omega_\Lambda(t)+\Omega_m(t)+\Omega_q(t)=1$
follows if conformal Eq.(\ref{MFeq}) is divided by 
${\dot a}^2/a^2$, defining dimensionless Friedmann weights
$\Omega_\Lambda=\frac{2}{3}\frac{{\bar\Lambda}a^2}{{\dot a}^2}$,
$\Omega_m=\frac{2}{3}\frac{{\bar\tau}c^2{\bar\rho}a^2}{{\dot a}^2}$,
and acceleration weight $\Omega_q=\frac{{\ddot a}a}{{\dot a}^2}$  
\cite{NESM1,NESM2}.  Matter and radiation are combined in $\Omega_m$ 
here while Hubble function $H(t)=\frac{{\dot a}}{a}(t)\sim [T^{-1}]=
h(t)H_0$  for Hubble constant $H_0=H(t_0)$. Setting $a(t_0)=1$,
$h(t_0)=\frac{{\dot a}}{a}(t_0)=1$ in Hubble units of time $1/H_0$,
length $c/H_0$, and acceleration $H_0^2c/H_0=cH_0$.
Evaluated at time $t$, Ricci scalar $R(t)=
 \frac{{\ddot a}}{a}+\frac{{\dot a}^2}{a^2}+\frac{k}{{\dot a}^2}$.
\par For $k=0$ the standard Friedmann equation,
divided by $\frac{{\dot a}}{a}$, produces dimensionless sum rule
$\Omega_\Lambda(t)+\Omega_m(t)=1$. $\Omega_m=1-\Omega_\Lambda$
requires mass density far greater than observed baryonic mass. 
This has been considered to be a strong argument for dark matter.
Omitting $\Omega_m$ completely, with $k=0$, conformal sum rule 
$\Omega_\Lambda(t)+\Omega_q(t)=1$ fits observed data accurately for 
redshifts $z\leq 1$ (7.33Gyr) \cite{NESM1,NESM2,NESM5}.
This eliminates any need for dark matter to explain Hubble expansion. 
\begin{table}[h]
\caption{Scaled luminosity distance fit to Hubble data} \label{TabA1}
\begin{tabular}{lcccc}
 &                 &          &Theory     &Observed    \\
z& $\Omega_\Lambda$&$\Omega_q$&$H_0d_L/c$Eq.(\ref{Chi_z})
 &$H_0d_L/c$\cite{MAN03}\\ \hline
0.0& 0.732& 0.268& 0.0000& 0.0000\\
0.2& 0.578& 0.422& 0.2254& 0.2265\\
0.4& 0.490& 0.510& 0.5013& 0.5039\\
0.6& 0.434& 0.566& 0.8267& 0.8297\\
0.8& 0.393& 0.607& 1.2003& 1.2026\\
1.0& 0.363& 0.637& 1.6209& 1.6216
\end{tabular}
\end{table}
Luminosity distance $d_L(z)=(1+z)\chi(z)$, for $\Omega_k=0$, is shown
in Table(\ref{TabA1}) for $\alpha=\Omega_\Lambda(t_0)=0.732$,
where\cite{NESM2}
\begin{eqnarray} \label{Chi_z}
\chi(z)=\int_{t(z)}^{t_0}\frac{dt}{a(t)}=
 \int_z^0 dz(1+z)\frac{dt}{dz}=
\nonumber\\
 \int_0^z\frac{dz}{\sqrt{2\alpha\ln(1+z)+1}}.
\end{eqnarray}
Observed redshifts have been fitted to an analytic 
function\cite{MAN03} with
statistical accuracy comparable to the best standard $\Lambda$CDM fit,
with $\Omega_m=0$. Table(\ref{TabA1}) compares CHM $d_L(z)$ to this
Mannheim function.  Because ${\bar\tau}$ is negative\cite{MAN06,NESM1},
cosmic acceleration $\Omega_q$ remains positive (centrifugal) back to
the earliest time\cite{NESM1}.
\par Conformal Friedmann Eq.(\ref{MFeq})\cite{NESM1,NESM2} determines 
cosmic acceleration weight $\Omega_q$. With both weight parameters 
$\Omega_k$ and $\Omega_m$ set to zero, Eq.(\ref{MFeq}) fits scaled 
Hubble function $h(t)=H(t)/H_0$ for redshifts $z\leq 1$,
Table(\ref{TabA1}), as accurately as standard LCDM, with only one 
free constant.  This determines Friedmann weights, at present time 
$t_0$, $\Omega_\Lambda=0.732, \Omega_q=0.268$ \cite{NESM1}.
Hubble constant $H(t_0)=H_0=2.197\times 10^{-18}/s$ \cite{PLC15}
is independent of these data.
The dimensionless sum rule\cite{NESM1} with $\Omega_k=0$ determines
$\Omega_q(\rho_m)=1-\Omega_\Lambda-\Omega_m$ in the cosmic background,  
and $\Omega_q(0)=1-\Omega_\Lambda$ in the depleted halo\cite{NESM3}.   

\section{Conformal gravity}
\par The unique Lagrangian density ${\cal L}_g$ of conformal gravity,
constructed from the conformal Weyl tensor \cite{WEY18,MAK89,MAN06}, 
determines Schwarzschild gravitational potential
\begin{eqnarray} 
B(r)=-2\beta/r+\alpha+\gamma r-\kappa r^2, 
\end{eqnarray}
valid outside a spherically symmetric
mass/energy source density \cite{MAK89,MAN91}.
Classical gravitation is retained at subgalactic distances
by setting $\beta=Gm/c^2$ for a spherical source of mass $m$.
CG adds two constants of integration to the 
classical external potential: nonclassical radial acceleration 
$\gamma$ and halo cutoff parameter $\kappa$ \cite{NESM3}, with
negligible effects at subgalactic distances\cite{MAN06}.
\par From the CHM, observed nonclassical gravitational acceleration 
$\half\gamma c^2$ in the halo is proportional to 
$\Delta\Omega_q=\Omega_q(0)-\Omega_q(\rho_m)=
\Omega_m(\rho_m)$\cite{NESM3}, where, given $\rho_m$ and $H_0$, 
$\Omega_m(\rho_m)=
\frac{2}{3}\frac{{\bar\tau}c^2\rho_m}{H^2_0}$\cite{NESM1}. 
Thus the depleted halo model determines constant $\gamma$ from
uniform universal cosmic baryonic mass density $\rho_m/c^2$,
which includes radiation energy density here.
\par Converted from Hubble units, this implies centripetal
acceleration $\half\gamma c^2=-cH_0\Omega_m(\rho_m)$\cite{NESM3}.
Positive $\rho_m$ implies $\Omega_m<0$ because coefficient 
${\bar\tau}<0$ \cite{MAN06,NESM1}. Hence 
$\Delta\Omega_q=\Omega_m<0$ is consistent with nonclassical
centripetal acceleration $\half\gamma c^2$, confirmed by inward
deflection of photon geodesics observed in gravitational
lensing \cite{NESM3}.  This logic is equivalent to requiring 
radial acceleration to be continuous across halo boundary $r_H$:
\begin{eqnarray}
\half\gamma_H c^2-cH_0\Omega_q(0)=-cH_0\Omega_q(\rho_m).
\end{eqnarray}
Signs here follow from the definition of $\Omega_q$ as centrifugal
acceleration weight.
\par Given mass/energy source density $f(r)$ enclosed within $\R$,
the field equation in the ES metric is \cite{MAK89,MAN91}
\begin{eqnarray} \label{Beq}
 \dar^4(rB(r))=rf(r), 
\end{eqnarray}
for $f(r)\sim(\Theta_0^0-\Theta_r^r)_m$ determined by source 
energy-momentum tensor $\Theta^{\mu\nu}_m$\cite{MAN06}.
\par
For constants related by $\alpha^2=1-6\beta\gamma$ \cite{MAN91},
\begin{eqnarray}
 y_0(r)=rB(r)=-2\beta+\alpha r+\gamma r^2-\kappa r^3 ,
\end{eqnarray}
is a solution of the tensorial field equation for
source-free $r\geq\R$ \cite{MAK89,MAN91}. 
Derivative functions $y_i(r) =\partial^i_r(rB(r)),0\leq i\leq3$
satisfy differential equations 
\begin{eqnarray}
 \dar y_i=y_{i+1},0\leq i\leq2, \nonumber\\
 \dar y_3=rf(r) .
\end{eqnarray}
The general solution, for
independent constants $c_i=y_i(0)$, determines coefficients
$\beta,\alpha,\gamma,\kappa$ such that at endpoint $\R$
\begin{eqnarray}
 y_0(\R)=-2\beta+\alpha \R+\gamma \R^2-\kappa \R^3,\nonumber\\
 y_1(\R)=\alpha+2\gamma \R-3\kappa \R^2,\nonumber\\
 y_2(\R)=2\gamma-6\kappa \R,\nonumber\\
 y_3(\R)=-6\kappa .
\end{eqnarray}
Gravitational potential $B(r)$ is required to be
differentiable and free of singularities. 
$c_0=0$ prevents a singularity at the origin.  Values 
of $c_1,c_2,c_3$ can be chosen to match outer boundary conditions
$\alpha=1,\gamma=0,\kappa=0$ at $r=\R$. 
This determines parameter $\beta$.
Specific values of $\gamma$ and $\kappa$, consistent with Hubble 
expansion and the observed galactic dark halo \cite{NESM1,NESM3}, 
are fitted by adjusting $c_1,c_2,c_3$, subject to 
$c_0=0,\alpha^2=1-6\beta\gamma$.
\par A particular solution for $B(r)$ \cite{MAK89,MAN91}, assumed by 
subsequent authors, derives an integral for $\gamma$ that vanishes for 
residual source density ${\hat\rho}$.  This is replaced here by an 
alternative solution for which $\gamma$ is a free parameter\cite{NESM5}. 
\par For a single spherical solar mass isolated in a galactic halo,
mean internal mass density ${\bar\rho}_\odot$ within $r_\odot$
determines an exact solution of the conformal Higgs gravitational
equation, giving internal acceleration 
$\Omega_q({\bar\rho}_\odot)$.
\par Given $\gamma$ outside $r_\odot$,
continuous acceleration across boundary $r_\odot$,
\begin{eqnarray}  
\half\gamma_{\odot,in}c^2-cH_0\Omega_q({\bar\rho}_\odot)=
 \half\gamma c^2-cH_0\Omega_q(0),
\end{eqnarray}
determines constant $\gamma_{\odot,in}$ valid inside $r_\odot$.
$\gamma_{\odot,in}$ is determined by local mean source density
${\bar\rho}_\odot$. $\gamma$ in the halo is not changed.  Its value is 
a constant of integration that cannot vary in the source-free 
halo\cite{NESM3,NESM5}.  Hence there is no way to determine
a mass-dependent increment to $\gamma$.  This replaces the usually 
assumed $\gamma=\gamma_0+N^*\gamma^*$ by $\gamma=\gamma_H$, 
determined at halo boundary $r_H$.
\par Anomalous rotation velocities for 138 galaxies have been fitted
to Eq.(\ref{Brfn}) using four assumed universal
parameters $\beta^*,\gamma^*,\gamma_0,\kappa$
\cite{MAN06,MAN12,MAO11,OAM15} such that
$\beta=N^*\beta^*=GM/c^2,\gamma=\gamma_0+N^*\gamma^*$.
$N^*$ is galactic baryonic mass $M$ in solar mass units.
Inferred parameter values \cite{MAN06,MAO12}, 
\begin{eqnarray}
\beta^*=1.475\times 10^3 m,
\gamma_0=3.06\times 10^{-28}/m, \nonumber\\
\gamma^*=5.42\times 10^{-39}/m,
\kappa  =9.54\times 10^{-50}/m^2,
\end{eqnarray}
fit conformal gravity to galactic rotation velocities.
The depleted halo model removes the galactic mass dependence of
nonclassical acceleration parameter $\gamma$. For our Milky Way galaxy,
with $N^*=6.07\times 10^{10}$ \cite{MCG08,OAM15},
implied $\gamma_H= \gamma_0+N^*\gamma^*=6.35\times 10^{-28}/m$.
\revision{
Parameter $\kappa$ provides a cutoff of modified radial acceleration
at a large halo radius\cite{NESM5}.
}

\section{Reconciliation of the two distinct gravitational models}
\par CG and CHM must be consistent for an isolated galaxy and its
dark halo, observed by gravitational lensing.  CG is valid for anomalous
outer galactic rotation velocities in the static spherical Schwarzschild
metric, solving a differential equation for Schwarzschild gravitational
potential $B(r)$ \cite{MAK89,MAN06}.  The CHM is valid for cosmic
Hubble expansion in the uniform, isotropic FLRW metric, solving a
differential equation for Friedmann scale factor $a(t)$ \cite{NESM1}.
Concurrent validity is achieved by introducing a common hybrid
metric.  The two resulting gravitational equations are decoupled by
separating mass/energy source density $\rho$ into uniform isotropic
mean density ${\bar\rho}$ and residual ${\hat\rho}=\rho-{\bar\rho}$,
which extends only to galactic radius $r_G$ and integrates to zero
over the defining volume.
\par The conformally invariant action integral of conformal gravity
is defined by a quadratic contraction of the Weyl tensor, simplified
by removing a 4-divergence \cite{MAN06,MAN12}.
For uniform density ${\bar\rho}$ the Weyl tensor vanishes identically, 
so that $X_g^{\mu\nu}\equiv 0$ for a uniform, isotropic cosmos. 
\par Observed excessive galactic rotational velocities have been
studied and parametrized using conformal Weyl Lagrangian density 
${\cal L}_g$ \cite{MAN06,MAN12}. The generalized Einstein 
equation exactly cancels any vacuum energy density.  Hubble expansion 
has been parametrized using conformal Higgs scalar field Lagrangian
density ${\cal L}_\Phi$ \cite{NESM1}.  
\par Metric tensor $g_{\mu\nu}$ is determined by conformal field 
equations derived from ${\cal L}_g+{\cal L}_\Phi$ \cite{NESM3},
driven by energy-momentum tensor $\Theta_m^{\mu\nu}$, where
subscript $m$ refers to conventional matter and radiation.
The gravitational field equation within halo radius $r_H$ is
\begin{eqnarray}
X_g^{\mu\nu}+X_\Phi^{\mu\nu}=\half\Theta_m^{\mu\nu}.
\end{eqnarray}
Defining mean density ${\bar\rho}_G$ and
residual density ${\hat\rho}_G=\rho_G-{\bar\rho}_G$,
and assuming
$\Theta_m^{\mu\nu}(\rho)\simeq
 \Theta_m^{\mu\nu}({\bar\rho})+\Theta_m^{\mu\nu}({\hat\rho})$,
solutions for $r\leq r_G$ of the two equations
\begin{eqnarray} \label{Twoeqs}
X_g^{\mu\nu}=\half\Theta_m^{\mu\nu}({\hat\rho}_G),
X_\Phi^{\mu\nu}=\half\Theta_m^{\mu\nu}({\bar\rho}_G)
\end{eqnarray}
decouple and imply a solution of the full equation.
This removes any mean density source from the $X_g$ equation, leaving
only the residual density, which integrates to zero over a closed
volume and cancels vacuum energy.
\par The decoupled solutions require a composite hybrid metric such 
as \cite{NESM5}
\begin{eqnarray} \label{xmet}
ds^2=-B(r)dt^2+a^2(t)(\frac{dr^2}{B(r)}+r^2d\omega^2). 
\end{eqnarray}
Solutions in the two distinct primitive metrics are made 
consistent by fitting parameters to boundary conditions and setting 
cosmic curvature constant $k=0$, justified by currently observed data. 

\section{Baryonic Tully-Fisher and radial acceleration relations}
\par Static spherical geometry defines Schwarzschild
potential $B(r)$.  For a test particle in a stable exterior
circular orbit with velocity $v$ the centripetal acceleration is
$a=v^2(r)/r=\half B'(r)c^2$.
Newtonian $B(r)=1-2\beta/r$, where $\beta=GM/c^2$,
so that $a_N=\beta c^2/r^2=GM/r^2$.
\par CG adds nonclassical $\Delta a$ to $a_N$, so that orbital velocity
squared is the sum of $v^2(a_N;r)$ and $v^2(\Delta a;r)$, which
cross with equal and opposite slope at some $r=r_{TF}$.  This defines
a flat range of $v(r)$ centered at stationary point $r_{TF}$, without
constraining behavior at large $r$.
\par MOND\cite{MIL83,SAN10,FAM12} modifies the Newtonian force
law for acceleration below an empirical scale $a_0$.
Using $y=a_N/a_0$ as independent variable,
for assumed universal constant $a_0\simeq 10^{-10}m/s^2$,
MOND postulates an interpolation function $\nu(y)$ such that observed
radial acceleration $a=f(a_N)=a_N\nu(y)$.
A flat velocity range approached
asymptotically requires $a^2\to a_Na_0$ as $a_N\to0$.
For $a_N\gg a_0$, $\nu\to1$ and for $a_N\ll a_0$,
$\nu^2\to 1/y$.  This implies asymptotic limit $a^2\to a_0a_N$
for small $a_N$, which translates into an
asymptotically flat galactic velocity function $v(r)$ for
large orbital radius $r$\cite{MIL83}.
For $a_N\ll a_0$, MOND $v^4=a^2r^2\to GMa_0$, the empirical
baryonic Tully-Fisher relation\cite{TAF77,MCG05,MCG11,OCM18}.
\par In conformal gravity (CG), centripetal acceleration
$a=v^2/r$ determines exterior orbital velocity
$v^2/c^2=ra/c^2=\beta/r+\half\gamma r-\kappa r^2$,
compared with asymptotic $ra_N/c^2=\beta/r$.
If the asymptotic Newtonian function is valid at $r$ and
$2\kappa r/\gamma$ can be neglected, the slope of $v^2(r)$ vanishes at
$r^2_{TF}=2\beta/\gamma$.  This implies that
$v^4(r_{TF})/c^4=(\beta/r_{TF}+\half\gamma r_{TF})^2=
  2\beta\gamma$\cite{MAN97,OCM18}.
This is the Tully-Fisher relation, exact at stationary point
$r_{TF}$ of the $v(r)$ function.  Given $\beta=GM/c^2$,
$v^4=2GM\gamma c^2$, for relatively constant $v(r)$ centered
at $r_{TF}$.  
\par McGaugh et al\cite{MLS16} have recently shown for 153 disk
galaxies that observed radial acceleration $a$ is effectively a 
universal function of the expected classical Newtonian acceleration 
$a_N$, computed for the observed baryonic distribution. Galactic mass 
is determined directly by observation, removing uncertainty due to 
adjustment of mass-to-light ratios for individual 
galaxies in earlier studies.
The existence of such a universal correlation function,
$a(a_N)=a_N\nu(a_N/a_0)$ is a basic postulate of
MOND\cite{MIL83,FAM12}. 
\par CG implies a similar correlation function if 
nonclassical parameter $\gamma$ is mass-independent\cite{NESM7}.
Outside an assumed spherical source mass, conformal Schwarzschild
potential $B(r)$ determines circular geodesics such that
$v^2/c^2=ra/c^2=\half rB^\prime(r)=\beta/r+\half\gamma r-\kappa r^2$.
The Kepler formula is $ra_N/c^2=\beta/r$.
Well inside a galactic halo boundary, $2\kappa r/\gamma$ can be
neglected.  This defines correlation function $a(a_N)=a_N+\Delta a$
if $\Delta a=\half\gamma c^2$ is a universal constant \cite{NESM7}, 
which is implied by the depleted halo model\cite{NESM3}. This requires 
reconsideration of the definition $\gamma=\gamma_0+N^*\gamma^*$ used in 
fitting rotation data for 138 galaxies to CG \cite{MAN06,MAO12}. 
For comparison with CG for the Tully-Fisher relation,
CG would agree with MOND $v^4=GMa_0$ if $a_0=2\gamma c^2$\cite{MAN97}, 
for mass-independent $\gamma$.  CG $\gamma=6.35\times 10^{-28}/m$ 
implies MOND $a_0=1.14\times 10^{-10}m/s^2$.

\section{Higgs parameter $\lambda$}
\par The conformal scalar field equation, including
parametrized $\Delta{\cal L}_\Phi$, is \cite{MAN06,NESM1,NESM5}
\begin{eqnarray}\label{Phieq}
\frac{1}{\sqrt{-g}}\partial_\mu(\sqrt{-g}\partial^\mu\Phi)=
 -(\sxth R-w_0^2+2\lambda\Phi^\dag\Phi)\Phi.
\end{eqnarray}
For $k=0$, $\sxth R(t)=
 h^2(t)(1+\Omega_q(t))>h^2(t)\Omega_\Lambda=w(t)^2$,
where $h(t)={\dot a}/a$ in Hubble units\cite{NESM1}.
\par Ricci scalar $R$ introduces gravitational effects.
Time-dependent $R(t)=6(\xi_0(t)+\xi_1(t))$, where
$\xi_0(t)=\frac{\ddot a}{a}$ and
$\xi_1(t)=\frac{{\dot a}^2}{a^2}+\frac{k}{a^2}$\cite{NESM1}. 
For $h(t)={\dot a}/a$ and $k=0$, $\sxth R(t)=
 h^2(t)(2-\Omega_\Lambda(t)-\Omega_m(t))=
 h^2(t)(1+\Omega_q(t))>h^2(t)\Omega_\Lambda(t)))=w(t)^2$.
Hence $\zeta(t)=\sxth R(t)-w(t)^2>0$.
\par Only real-valued solution $\phi(t)$ is relevant in uniform,
isotropic geometry. The field equation is
\begin{eqnarray} \label{phieq}
\frac{{\ddot\phi}}{\phi}+3\frac{{\dot a}}{a}\frac{{\dot\phi}}{\phi}=
 -(\sxth R(t)-w(t)^2+2\lambda\phi^2).
\end{eqnarray}
\par Omitting $R$ and assuming constant $\lambda>0$, Higgs
solution $\phi_0^2=w_0^2/2\lambda$ \cite{HIG64} is exact.  
All time derivatives drop out. 
In the conformal scalar field equation,
cosmological time dependence of Ricci scalar $R(t)$,
determined by the CHM Friedmann cosmic evolution equation,
introduces nonvanishing time derivatives.
$w_0^2=\frac{{\dot\phi}^2}{\phi^2}$ from the Higgs field covariant 
derivative is consistent with constant $\lambda$.
Time-dependent terms in the scalar field equation can be included in
$w^2(t)=\frac{{\dot\phi}^2}{\phi^2}-\frac{{\ddot\phi}}{\phi}
-3h(t)\frac{{\dot\phi}}{\phi}$. 
For $\zeta(t)=\sxth R(t)-w^2(t)>0$, $\phi^2(t)=-\zeta(t)/2\lambda$
is an exact solution of Eq.(\ref{phieq}). 
$\zeta>0$ for computed $R(t)$\cite{NESM1} implies $\lambda<0$.
\par $\hbar\phi(t_0)=174GeV$\cite{AMS08}$=1.203\times 10^{44}\hbar H_0$. 
In Hubble units, for $\Omega_m=0$, $\zeta(t_0)=2\Omega_q(t_0)=0.536$.
For empirical $\phi(t)$, $\lambda(t)=\zeta/(-2\phi^2)$. Given 
$\zeta(t_0)$ and $\phi(t_0)$, 
dimensionless $\lambda(t_0)=-0.185\times 10^{-88}$ \cite{NESM5}.  

\section{$W_2$ particle and $Z_2$ resonance}
\revision{
\par Conformal theory obtains empirical parameters that preserve the 
standard electroweak model, but preclude a massive Higgs 
particle\cite{NESM5}.
}
The observed LHC 125GeV resonance\cite{DJV12,ATL12,CMS12} 
requires an alternative explanation. 
\par A model Hamiltonian matrix can be defined \cite{NESM4}
in which indices 0,1 refer respectively to bare neutral scalar states 
$WW=g_{\mu\nu}W_-^\mu W_+^\nu$, $ZZ=g_{\mu\nu}Z^{\mu*} Z^\nu$. Assumed 
diagonal elements are $H_{00}=2m_W=160GeV$, $H_{11}=2m_Z=182GeV$,
for empirical masses $m_W$ and $m_Z$.
Intermediate quark and lepton states define a large complementary matrix
${\tilde H}$ indexed by $i,j\neq 0,1$, with eigenvalues $\epsilon_i$, 
and off-diagonal elements ${\tilde A}_{i0},{\tilde A}_{i1}$.  
${\tilde H}$ determines energy-dependent increments in a
$2\times 2$ reduced matrix 
\begin{eqnarray}
 H_{ab}-\mu_{ab}=H_{ab}-\sum_{i \neq 0,1}
{\tilde A}^\dag_{ai}(\epsilon_i-\epsilon)^{-1}{\tilde A}_{ib} .
\end{eqnarray}
\par $H_{01}-\mu_{01}=(WW|H_{red}|ZZ)$ corresponds to Feynman diagrams
for quark and lepton exchange.  The most massive and presumably most
strongly coupled intermediate field that interacts directly would be
tetraquark $T=t{\bar b}b{\bar t}$, whose mass is estimated 
as $\epsilon_T=350$GeV.  The very strong interaction between bare 
fields ZZ and WW is assumed to be dominated by tetraquark exchange.  
\par A simplified estimate of $W_2$ energy is obtained by restricting 
intermediate states to the three color-indexed tetraquark states
$T=t{\bar b}b{\bar t}$, and assuming elements 
${\tilde A}_{T0},{\tilde A}_{T1}$ of equal magnitude $\alpha/\sqrt{3}$.
For the reduced $2\times2$ matrix, matrix increments
$\mu_{ab}\simeq\mu(\epsilon)=
\frac{\alpha^2}{\epsilon_T-\epsilon}$
are all defined by a single parameter $\alpha^2$.
Secular equation 
\begin{eqnarray}
 (2m_W-\mu(\epsilon)-\epsilon)(2m_Z-\mu(\epsilon)-\epsilon)=
\mu^2(\epsilon) 
\end{eqnarray}
is to be solved for two eigenvalues $\epsilon=E_0,E_1$.
\par It is found that identifying the model diboson $W_2$ with the
recently observed LHC 125GeV resonance\cite{DJV12,ATL12,CMS12}
confirms the empirical value of Higgs parameter $\lambda$.  In Higgs 
$V(\Phi^\dag\Phi)=-(w^2 -\lambda \Phi^\dag\Phi)\Phi^\dag\Phi$,
coefficient $w^2$ results from dressing the bare massless
Higgs scalar field by neutral gauge field $Z_\mu$, while coefficient 
$\lambda$ results from dressing by $W_2$ \cite{NESM4,NESM5}.
\par Setting $E_0=125GeV=0.8644\times 10^{44}\hbar H_0$ 
for the $W_2$ state, dominated by the bare 
$WW$ field, determines parameters $\alpha^2=4878$ GeV$^2$,
$\mu(E_0)=21.68$GeV and $\tan\theta_x=0.6138$.
Using $\alpha^2$ determined by $E_0$, the present model predicts 
$E_1=173$GeV, with $\mu(E_1)=27.62$GeV.  This higher eigenvalue 
is the energy of a resonance $Z_2$ dominated by the bare $ZZ$ field.
$Z_2$ decay into bare $WW$, two free charged gauge bosons, is allowed
by energy conservation, but not into bare $ZZ$.  Composite field $W_2$ 
cannot decay spontaneously into either $WW$ or $ZZ$.
\par Identifying $E_0$ with the observed 125GeV resonance, and using 
$g_w=0.6312$ and $g_z=0.7165$ with computed $\tan\theta_x=0.6138$,
the implied value of Higgs parameter 
$\lambda=-\frth g_w^2g_z^2
\sin2\theta_x(\frac{{\dot\phi}_0}{\phi_0})^2\hbar^2/m^2_{W_2}c^4
=-0.455\times10^{-88}$ is consistent with its empirical 
value $\lambda\simeq -10^{-88}$ \cite{NESM4,NESM5}.

\vfill\eject
\end{document}